\documentstyle[12pt]{article}

\catcode`\@=11
\long\def\@makefntext#1{ 
\protect\noindent \hbox to 3.2pt {\hskip-.9pt  
$^{{\ninerm\@thefnmark}}$\hfil}#1\hfill} 

\def\thefootnote{\fnsymbol{footnote}}
 \def\@makefnmark{\hbox to 0pt{$^{\@thefnmark}$\hss}}  
	
\def\ps@myheadings{\let\@mkboth\@gobbletwo
\def\@oddhead{\hbox{} 
\rightmark\hfil\ninerm\thepage}   
\def\@oddfoot{}\def\@evenhead{\ninerm\thepage\hfil 
\leftmark\hbox{}}\def\@evenfoot{}
\def\sectionmark##1{}\def\subsectionmark##1{}}

\begin{document}

\newcommand{\symbolfootnote}{\renewcommand{\thefootnote}
	{\fnsymbol{footnote}}}
\renewcommand{\thefootnote}{\fnsymbol{footnote}}
\newcommand{\alphfootnote}
	{\setcounter{footnote}{0}
	 \renewcommand{\thefootnote}{\sevenrm\alph{footnote}}}

\newcounter{sectionc}\newcounter{subsectionc}\newcounter{subsubsectionc}
\renewcommand{\section}[1] {\vspace{0.6cm}\addtocounter{sectionc}{1} 
\setcounter{subsectionc}{0}\setcounter{subsubsectionc}{0}\noindent 
	{\bf\thesectionc. #1}\par\vspace{0.4cm}}
\renewcommand{\subsection}[1] {\vspace{0.6cm}\addtocounter{subsectionc}{1} 
	\setcounter{subsubsectionc}{0}\noindent 
	{\it\thesectionc.\thesubsectionc. #1}\par\vspace{0.4cm}}
\renewcommand{\subsubsection}[1] {\vspace{0.6cm}\addtocounter{subsubsectionc}{1}
	\noindent {\rm\thesectionc.\thesubsectionc.\thesubsubsectionc. 
	#1}\par\vspace{0.4cm}}
\newcommand{\nonumsection}[1] {\vspace{0.6cm}\noindent{\bf #1}
	\par\vspace{0.4cm}}
					         
\newcounter{appendixc}
\newcounter{subappendixc}[appendixc]
\newcounter{subsubappendixc}[subappendixc]
\renewcommand{\thesubappendixc}{\Alph{appendixc}.\arabic{subappendixc}}
\renewcommand{\thesubsubappendixc}
	{\Alph{appendixc}.\arabic{subappendixc}.\arabic{subsubappendixc}}

\renewcommand{\appendix}[1] {\vspace{0.6cm}
        \refstepcounter{appendixc}
        \setcounter{figure}{0}
        \setcounter{table}{0}
        \setcounter{equation}{0}
        \renewcommand{\thefigure}{\Alph{appendixc}.\arabic{figure}}
        \renewcommand{\thetable}{\Alph{appendixc}.\arabic{table}}
        \renewcommand{\theappendixc}{\Alph{appendixc}}
        \renewcommand{\theequation}{\Alph{appendixc}.\arabic{equation}}
        \noindent{\bf Appendix \theappendixc #1}\par\vspace{0.4cm}}
\newcommand{\subappendix}[1] {\vspace{0.6cm}
        \refstepcounter{subappendixc}
        \noindent{\bf Appendix \thesubappendixc. #1}\par\vspace{0.4cm}}
\newcommand{\subsubappendix}[1] {\vspace{0.6cm}
        \refstepcounter{subsubappendixc}
        \noindent{\it Appendix \thesubsubappendixc. #1}
	\par\vspace{0.4cm}}

\def\abstracts#1{{
	\centering{\begin{minipage}{30pc}\tenrm\baselineskip=12pt\noindent
	\centerline{\tenrm ABSTRACT}\vspace{0.3cm}
	\parindent=0pt #1
	\end{minipage} }\par}} 

\newcommand{\bibit}{\it}
\newcommand{\bibbf}{\bf}
\renewenvironment{thebibliography}[1]
	{\begin{list}{\arabic{enumi}.}
	{\usecounter{enumi}\setlength{\parsep}{0pt}
\setlength{\leftmargin 1.25cm}{\rightmargin 0pt}
	 \setlength{\itemsep}{0pt} \settowidth
	{\labelwidth}{#1.}\sloppy}}{\end{list}}

\topsep=0in\parsep=0in\itemsep=0in
\parindent=1.5pc

\newcounter{itemlistc}
\newcounter{romanlistc}
\newcounter{alphlistc}
\newcounter{arabiclistc}
\newenvironment{itemlist}
    	{\setcounter{itemlistc}{0}
	 \begin{list}{$\bullet$}
	{\usecounter{itemlistc}
	 \setlength{\parsep}{0pt}
	 \setlength{\itemsep}{0pt}}}{\end{list}}

\newenvironment{romanlist}
	{\setcounter{romanlistc}{0}
	 \begin{list}{$($\roman{romanlistc}$)$}
	{\usecounter{romanlistc}
	 \setlength{\parsep}{0pt}
	 \setlength{\itemsep}{0pt}}}{\end{list}}

\newenvironment{alphlist}
	{\setcounter{alphlistc}{0}
	 \begin{list}{$($\alph{alphlistc}$)$}
	{\usecounter{alphlistc}
	 \setlength{\parsep}{0pt}
	 \setlength{\itemsep}{0pt}}}{\end{list}}

\newenvironment{arabiclist}
	{\setcounter{arabiclistc}{0}
	 \begin{list}{\arabic{arabiclistc}}
	{\usecounter{arabiclistc}
	 \setlength{\parsep}{0pt}
	 \setlength{\itemsep}{0pt}}}{\end{list}}

\newcommand{\fcaption}[1]{
        \refstepcounter{figure}
        \setbox\@tempboxa = \hbox{\tenrm Fig.~\thefigure. #1}
        \ifdim \wd\@tempboxa > 6in
           {\begin{center}
        \parbox{6in}{\tenrm\baselineskip=12pt Fig.~\thefigure. #1 }
            \end{center}}
        \else
             {\begin{center}
             {\tenrm Fig.~\thefigure. #1}
              \end{center}}
        \fi}

\newcommand{\tcaption}[1]{
        \refstepcounter{table}
        \setbox\@tempboxa = \hbox{\tenrm Table~\thetable. #1}
        \ifdim \wd\@tempboxa > 6in
           {\begin{center}
        \parbox{6in}{\tenrm\baselineskip=12pt Table~\thetable. #1 }
            \end{center}}
        \else
             {\begin{center}
             {\tenrm Table~\thetable. #1}
              \end{center}}
        \fi}

\def\@citex[#1]#2{\if@filesw\immediate\write\@auxout
	{\string\citation{#2}}\fi
\def\@citea{}\@cite{\@for\@citeb:=#2\do
	{\@citea\def\@citea{,}\@ifundefined
	{b@\@citeb}{{\bf ?}\@warning
	{Citation `\@citeb' on page \thepage \space undefined}}
	{\csname b@\@citeb\endcsname}}}{#1}}

\newif\if@cghi
\def\cite{\@cghitrue\@ifnextchar [{\@tempswatrue
	\@citex}{\@tempswafalse\@citex[]}}
\def\citelow{\@cghifalse\@ifnextchar [{\@tempswatrue
	\@citex}{\@tempswafalse\@citex[]}}
\def\@cite#1#2{{$\null^{#1}$\if@tempswa\typeout
	{IJCGA warning: optional citation argument 
	ignored: `#2'} \fi}}
\newcommand{\citeup}{\cite}

\def\fnm#1{$^{\mbox{\scriptsize #1}}$}
\def\fnt#1#2{\footnotetext{\kern-.3em
	{$^{\mbox{\sevenrm #1}}$}{#2}}}

\font\twelvebf=cmbx10 scaled\magstep 1
\font\twelverm=cmr10 scaled\magstep 1
\font\twelveit=cmti10 scaled\magstep 1
\font\elevenbfit=cmbxti10 scaled\magstephalf
\font\elevenbf=cmbx10 scaled\magstephalf
\font\elevenrm=cmr10 scaled\magstephalf
\font\elevenit=cmti10 scaled\magstephalf
\font\bfit=cmbxti10
\font\tenbf=cmbx10
\font\tenrm=cmr10
\font\tenit=cmti10
\font\ninebf=cmbx9
\font\ninerm=cmr9
\font\nineit=cmti9
\font\eightbf=cmbx8
\font\eightrm=cmr8
\font\eightit=cmti8

\newcommand{\RMA}{{\cal R}}     
\newcommand{\Acr}{{\cal A}}     
\newcommand{\ALG}{{\cal A}}     
\newcommand{\HH}{{\cal H}}      
\newcommand{\CC}{{\rm C}}       
\newcommand{\RR}{{\rm R}}       
\newcommand{\Id}{{\rm I}}       
\newcommand{\PP}{{\rm P}}       
\newcommand{\hh}{h}             
\newcommand{\HAM}{H}            
\newcommand{\ham}{h}            
\newcommand{\VV}{V}             
\newcommand{\UU}{U}             
\newcommand{\pp}{p}             
\newcommand{\PMO}{P}            
\newcommand{\al}{\alpha}
\newcommand{\bet}{\beta}
\newcommand{\ga}{\gamma}
\newcommand{\Ga}{\Gamma}
\newcommand{\dl}{\delta}
\newcommand{\ep}{\epsilon}
\newcommand{\Dl}{\Delta}
\newcommand{\si}{\sigma}
\newcommand{\ka}{\kappa}
\newcommand{\lam}{\lambda}
\newcommand{\om}{\omega}
\newcommand{\Om}{\Omega}
\newcommand{\veps}{\varepsilon}
\newcommand{\vph}{\varphi}


\hfil{AS-ITP-96-33}
\vspace{0.9cm}
\centerline{\tenbf Super-Yangian $Y(gl(1|1))$ and Its Oscillator Realization
}
\vspace{0.8cm}
\centerline{\tenrm Guo-xin JU}
\baselineskip=13pt
\centerline{\tenit Institute of Theorectical Physics,Academia Sinica}
\baselineskip=12pt
\centerline{\tenit Beijing 100080,China}
\baselineskip=12pt
\centerline{\tenrm and}
\baselineskip=12pt
\centerline{\tenit Physics Department,Henan Normal University}
\baselineskip=12pt
\centerline{\tenit Xinxiang,Henan Province 453002,China}
\vspace{0.3cm}
\centerline{\tenrm Jin-fang CAI, Han-ying GUO , Ke WU}
\baselineskip=13pt
\centerline{\tenit Institute of Theorectical Physics,Academia Sinica}
\baselineskip=12pt
\centerline{\tenit Beijing 100080,China}
\vspace{0.3cm}
\centerline{\tenrm and}
\vspace{0.3cm}
\centerline{\tenrm Shi-kun WANG}
\baselineskip=13pt
\centerline{\tenit Institute of Applied Mathematics,Academia Sinica}
\baselineskip=12pt
\centerline{\tenit Beijing 100080,China}
\vspace{0.9cm}

\abstracts{On the basis of graded $RTT$ formalism,the defining relation of the 
super-Yangian $Y(gl(1|1))$ is derived and its oscillator realization is 
constructed.}

\vfil
\twelverm   
\baselineskip=14pt
\section{Introduction }
\vspace*{-0.35cm}

Yangian $Y(g)$ of a simple Lie algebra $g$,first introduced by V. G.Drinfeld 
\cite{Dri}
,is a deformation of the universal enveloping algebra U(g[t]) of a current 
algebra g[t].It is a kind of Hopf algebra and the tensor products of its 
finite-dimensional representations produce rational solutions of the quantum 
Yang-Baxter equation(QYBE).\\

In the last decade,Yangians associated with simple Lie algebra have been 
systematically studied both in mathematics and physics \cite{Mol/Na},and have 
many applications in such theoretical physics as quantum field theory and 
statistical
mechanics.Yangian structure is the underlying symmetry of many types of 
integrable models.For example,
1-D Hubbard model on the infinite chain \cite{Ugl/Kor},the
Haldane-Shastry model and the 
Polychronakos-Frahm model \cite{Hik} have Yangian symmetry;in the massive 2-D 
quantum field theory,a infinite-dimensional symmetry
generated by nonlocal concerved currents is connected to the Yangian 
\cite{Lec/Smi}.\\

As generalizations of Yangians of simple Lie algebras,the Yangians
associated with the simple Lie superalgebra,
which we will call super-Yangian in this letter,also need to be 
studied.Actually,some structural features of super-Yangian
have been investegated by M. Nazarov \cite{Naz} and R. B. Zhang 
\cite{Zha,Zha1}.In ref 6,the quantum determinant of the super-Yangian 
$Y(gl(m|n))$ is discribed,
while in refs.7 and 8 the super-Yangian $Y_{q}(gl(m|n))$ asssociated with the 
Perk-Schultz $\RMA$ matrix is constructed,its structural properties and the 
relationship between its central elements and the Casimir operators of quantum 
supergroup $U_{q}(gl(m|n))$ is discussed,in particular,the classification of the 
finite-dimensional irreducible representations of the super-Yangians 
$Y(gl(1|1))$ and $Y(gl(m|n))$ is given.\\

In this letter,on the basis of the graded $RTT$ formalism,we derived the 
defining relations of the super-Yangian for the Lie superalgebra $gl(1|1)$ and 
give its oscillator realization.In section 2,we briefly review the graded $RTT$ 
formalism and the 
corresponding graded Yang-Baxter equation(GYBE). In section 3,we give the 
algebraic relation that super-Yangian $Y(gl(1|1))$ satisfies and construct its 
oscillator realization.Finally,we make some remarkes and discussions.

\section{Graded $RTT$ Relation and GYBE}
\vspace*{-0.35cm}
In the supersymmetric case,space is graded and the  tensor product has the 
following property

 $$
(A\otimes B)(C\otimes D)=(-1)^{p(B)p(C)}AC\otimes BD
\eqno{(2.1)}
$$
\noindent
where p(A) denotes the degree of A.Now the graded $RTT$ relation with the 
spectral parameters takes the form \cite{Liao/Song}

$$
\RMA _{12}(u-v)T_1 (u) \eta _{12} T_2 (v) \eta _{12} = \eta _{12} T_2 (v) \eta 
_{12} T_1 (u) \RMA _{12}(u-v) 
\eqno{(2.2a)}
$$
where $T_1 (u)=T(u)\otimes 1$ and $T_2 (u)=1\otimes T(u)$ and $(\eta  
_{12})_{ab,cd}=(-1)^{p(a)p(b)} \delta _{ac} \delta _{bd}$,and GYBE with spectral 
parameters reads as \cite{Liao/Song}

$$ 
\eta _{12} \RMA _{12}(u) \eta _{13}\RMA _{13}(u+v) \eta _{23}\RMA _{23}(v) 
= \eta _{23}\RMA _{23}(v) \eta _{13}\RMA _{13}(u+v) \eta _{12}\RMA _{12}(u), 
\eqno{(2.3a)}
$$

Considering the charge conservation conditions for the $\RMA _{ab,cd}$,i.e.
$$
\RMA _{ab,cd}=0 
\hskip 1cm 
unless 
\hskip 0.5cm
a+b=c+d
\eqno{(2.4)}
$$
we can write eqs. (2.2a) and (2.3a) in the component forms as follows:
 
$$
\begin{array}{l}
(-1)^{p(e)(p(d)+p(f))}\RMA _{12}(u-v)_{ab,cd} T(u)_{ce} T(v)_{df} =\\

(-1)^{p(a)(p(d)+p(b))} T(v)_{be}  T(u)_{ad} \RMA _{12}(u-v)_{cd,ef} 
\end{array} 
\eqno{(2.2b)}
$$

$$
\begin{array}{l}
(-1)^{p(d)(p(b)+p(e))}\RMA (u)_{ab,cd} \RMA (u+v)_{ce,fg} \RMA (v) _{dh,ij}
=\\

(-1)^{p(d)(p(h)+p(j))}\RMA (v)_{be,dh} \RMA (u+v)_{ah,cj} \RMA (u)_{cd,fi} 
\end{array} 
\eqno{(2.3b)}
$$
where the repeated indices is understood to take summation.
Note that, in Eqs. (2.2) and (2.3) the grading property is taken into account
by introducing the factor $\eta $.If we set $\eta =1$,then Eqs. (2.2) and (2.3) 
reduce to the usual $RTT$ relation and YBE respectively.

\section{Super-Yangian $Y(gl(1|1))$ and Its Oscillator Realization  }
\vspace*{-0.35cm}
It is well known  that

$$
\RMA _{12}(u)=u+{\cal P}_{12}
\eqno{(3.1)}
$$
satifies GYBE (2.3),where
$$
{\cal P}_{12} =\eta _{12} P_{12},
\eqno{(3.2)}
$$
P stands for the usual permutation operator,i.e.$P(u\otimes v)=v\otimes 
u$.Substituting eq (3.1) into eq (2.2) and introducing the notation

$$
[T(u)_{ab}, T(v)_{cd}\}=T(u)_{ab}T(v)_{cd}-(-1)^{(p(a)+p(b))(p(c)+p(d))}   
T(v)_{cd}T(u)_{ab}
\eqno{(3.3)}
$$
we obtain the following relations:

$$
(u-v)[T(u)_{ab},T(v)_{cd}\}+(-1)^{p(a)p(c)+p(a)p(b)+p(b)p(c)} 
(T(u)_{cb}T(v)_{ad}-T(v)_{cb}T(u)_{ad})=0
\eqno{(3.4)}
$$

Let
$$
T(u)_{ab}=\sum^{\infty}_{n=0}u^{-n}T^{(n)}_{ab}
\eqno{(3.5)}
$$
then from eq (3.4),we have

$$
[T^{(0)}_{ab}, T^{(n)}_{cd}\}=0
\eqno{(3.6)}
$$

$$
[T^{(n+1)}_{ab},T^{(m)}_{cd}\}-[T^{(n)}_{ab},T^{(m+1)}_{cd}\}\\
+(-1)^{p(a)p(c)+p(a)p(b)+p(b)p(c)}(T^{(n)}_{cb}T^{(m)}_{ad}-T^{(m)}_{cb}T^{(n)}_
{ad})=0 
\eqno{(3.7a)}
 $$

Similar to the discussion for the Yangian \cite{Mol/Na},eq (3.7a) can be 
rewritten into the following equivelent form:

$$
[T^{(n)}_{ab},T^{(m)}_{cd}\} 
=(-1)^{1+p(a)p(c)+p(a)p(b)+p(b)p(c)}\sum^{min(n,m)-1}_{i=0}(T^{(i)}_{cb}T^{(m+n-
i-1)}_{ad}-T^{(m+n-i-1)}_{cb}T^{(i)}_{ad})
\eqno{(3.7b)}
$$
In particular,for the case of $a=c,b=d$ in the above equation,we have

$$
[T^{(n)}_{ab},T^{(m)}_{ab}\} =(-1)^{1+p(a)p(a)} 
\sum^{min(n,m)-1}_{i=0}[T^{(i)}_{ab},T^{(m+n-i-1)}_{ab}]
\eqno{(3.8)}
$$
this shows that $T^{(n)}_{ab}$,with $a\not= b$ and different n $(n>1)$ will 
neither commute nor anticommute.\\

From eq (3.7b),we know that the following
property holds

$$
[T^{(n)}_{ab},T^{(m)}_{cd}\}=[T^{(m)}_{ab},T^{(n)}_{cd}\}\   \ (n,m\geq 1)
\eqno{(3.9)}
$$
Here we notice that in the non-graded case,eq (3.9) will give the relation

$$
[T^{(n)}_{ab},T^{(m)}_{cd}]=0\   \ (n,m\geq 1)
\eqno{(3.10)}
$$

For the case of superalgebra $gl(1|1)$,$a=1,2$ and ${\cal P}$ takes the form
$$
{\cal P}_{12}=\left[
\begin{array}{cccc}
1&0&0&0\\
0&0&1&0\\
0&1&0&0\\
0&0&0&-1
\end{array}
\right]
$$
$T(u)$ is a $2\times 2$ matrix.Because of the relation (3.6),we can choose 
$T^{(0)}$ to be of the form

$$
T^{(0)}=\left[
\begin{array}{cc}
1&0\\0&1
\end{array}\right]\ \
\eqno{(3.11)}
$$
up to a constant factor.Here we should stress that (3.11) is only a choice,which 
different from the non-graded case in that there it is the result of the Schur's 
lemma \cite{Fra/So}.From eq. (3.5),we see that eq.(3.11) is equivalent to impose 
asymptotic condition $T(u)\rightarrow 1$ for $u\rightarrow \infty$. 
With eqs (3.6) (3.7) and (3.11), we obtain the following relations:
$$
\left \{
\begin{array}{l}
[T^{(n)}_3 ,T^{(1)}_{12}]=[T^{(1)}_3 ,T^{(n)}_{12}]=0,\\

[T^{(n)}_3 ,T^{(1)}_{21}]=[T^{(1)}_3 ,T^{(n)}_{21}]=0,\\

[T^{(n)}_0 ,T^{(1)}_{12}]=[T^{(1)}_0 ,T^{(n)}_{12}]=-2 T^{(n)}_{12}, \hskip 
0.5cm
\mbox{(for any n)}\\

[T^{(n)}_0 ,T^{(1)}_{21}]=[T^{(1)}_0 ,T^{(n)}_{21}]=2   T^{(n)}_{21},\\

\{T^{(n)}_{12},T^{(1)}_{21}\}=-T^{(n)}_3 
\end{array}
\right.
\eqno{(3.12)}
$$

$$
\left \{
\begin{array}{l} 
[T^{(2)}_0 , T^{(2)}_3 ]+2  (T^{(1)}_{21}
T^{(2)}_{12}-T^{(2)}_{21}T^{(1)}_{12})=0\\
{[T^{(n)}_3 , T^{(2)}_{12}]}-T^{(1)}_{12}
T^{(n)}_3 +T^{(n)}_{12}T^{(1)}_3 =0\ \ (n\geq 1)\\
{[T^{(n)}_3 , T^{(2)}_{21}]}+T^{(1)}_{21}T^{(n)}_3 -T^{(n)}_{21}
T^{(1)}_3 =0\ \ (n\geq 1)
\end{array}
\right.
\eqno{(3.13)}
$$
and

$$
\left \{
\begin{array}{l}
-T^{(n+1)}_{12}=\{[T^{(n)}_0 , T^{(2)}_{12}]+
T^{(n)}_{12}T^{(1)}_0 -T^{(1)}_{12}T^{(n)}_0 \}/2\\

T^{(n+1)}_{21}=\{[T^{(n)}_0 ,
T^{(2)}_{21}]+T^{(1)}_{21}
T^{(n)}_0 -T^{(n)}_{21}T^{(1)}_0 \}/2\\

T^{(n+1)}_3 =-\{T^{(n)}_{12}, T^{(2)}_{21}\}+T^{(1)}_{22}
T^{(n)}_{11}-T^{(n)}_{22}T^{(1)}_{11},\ (n\geq 2)
\end{array}
\right.
\eqno{(3.14)}
$$
where

$$
T^{(n)}_3 =T^{(n)}_{22}-  
T^{(n)}_{11},
T^{(n)}_0 =T^{(n)}_{22}+   T^{(n)}_{11}
\eqno{(3.15)}
$$

From  the recurrence relations (3.14),we see that only $T^{(1)}_{ab}$, 
$T^{(2)}_{ab}$ are basic operators.\\
Now,if we make the following correspondence.

$$
\left \{
\begin{array}{ll}
T^{(1)}_3 =-\ga _0 z_0,  & T^{(2)}_3 =-\ga _1 z_1\\

T^{(1)}_{12}=\al _0 e_0,  &   T^{(2)}_{12}=\al _1 e_1\\

T^{(1)}_{21}=\bet _0 f_0,  &   T^{(2)}_{21}=\bet _1 f_1\\

T^{(1)}_0 =-2   h_0,&  T^{(2)}_0 =\dl h_1
\end{array}
\right.
\eqno{(3.16)}
$$
and take the choice

$$
\al _0 \bet _0=\ga _0, \al _0 \bet _1=\al _1 \bet _0=\ga _1, 
\al _0 \dl =-2   \al _1
\eqno{(3.17)}
$$
then from eqs (3.12)-(3.14) we obtain following algebraic relations

$$
\left \{
\begin{array}{ll}
e_0^2=f_0^2=0,[h_0,e_0]=e_0 ,&[h_0,f_0]=-f_0,\\

[z_0,e_0]=[z_0,f_0]=[h_0,z_0]=0, & \{e_0,f_0\}=z_0
\end{array}
\right.
\eqno{(3.18)}
$$
and

$$
\left \{
\begin{array}{ll}
[z_1,e_0]=[z_1,f_0]=[z_1,z_0]=[z_1,h_0]=0&\\

[f_1,z_0]=0,  &       [f_1,h_0]=f_1\\

\{f_1,e_0\}=z_1,  &   \{f_1,f_0\}=0\\

\{e_1,e_0\}=0 ,   &   \{e_1,f_0\}=z_1\\

[e_1,z_0]=0,    &     [e_1,h_0]=-e_1\\

[h_1,z_0]=[h_1,h_0]=0&\\

[h_1,e_0]=e_1,    &    [h_1,f_0]=-f_1
\end{array}
\right.
\eqno{(3.19)}
$$
The eq (3.18) is just the defining relation of the Lie superalgebra 
$gl(1|1)$.Taking the correspondence

$$
z_0\longrightarrow N+M,e_0\longrightarrow x, f_0\longrightarrow 
y,h_0\longrightarrow N
\eqno{(3.20)}
$$
eq (3.18) will give the same result as that of Liao and Song \cite{Liao/Song} in 
the limit $q\longrightarrow 1$.Eq (3.19) shows that $e_1 ,f_1,h_1,z_1$ form a 
representation of eq (3.18). $e_i ,f_i,h_i,z_i(i=0,1)$ also satisfies Serre 
relations:

$$
\begin{array}{l}
[z_1,\{ e_1,f_1 \} ]=C_0 z_0 (f_0 e_1 -f_1 e_0 )\\

2\{ e_1,[h_1,e_1 ]\} + C_0 [e_0 ,e_1] +2C_1 [h_1,e_1 ] e_0=0\\

2\{ f_1,[h_1,f_1 ]\} + C_0 [f_0 ,f_1] +2C_1 [h_1,f_1 ] f_0=0\\

\{ e_1,[z_1 ,e_1 ]\} + C_0 e_1 e_0 z_0 =0\\

\{ f_1,[z_1 ,f_1 ]\} + C_0 f_1 f_0 z_0 =0\\

[e_1,\{ e_1, f_1 \} ]+[ z_1,[h_1 ,e_1 ]]=C_1 ([e_1 ,h_1 ]z_0 +z_1 e_1) +C_0 e_0 
(f_0 e_1 -f_1 e_0 )\\

[f_1,\{ e_1, f_1 \} ]-[ z_1,[h_1 ,f_1 ]]=C_1 ([f_1 ,h_1 ]z_0 -z_1 f_1) +C_0 f_0 
(f_0 e_1 -f_1 e_0 )\\

[h_1,\{ e_1, f_1 \} ]-C_1 (f_0 [h_1 ,e_1] + [h_1, f_1 ]e_0 )-C_0 (f_0 e_1 -f_1 
e_0 )=0.
\end{array}
\eqno{(3.21)}
$$
where 

$$
C_0 =\ga _0 /\al_1 \bet _1, C_1 =\ga _1 /\al_1 \bet _1
$$
The operators $\{e_i,f_i,h_i, z_i\}_{i=0,1}$ and the relation (3.18) (3.19) and 
(3.21) constitute an infinite-dimensional algebra called super-Yangian of the 
Lie superalgebra $gl(1|1)$ and denoted by $Y(gl(1|1))$.$Y(gl(1|1))$ is a Hopf 
algebra with the comultiplication $\Dl$,co-unit $\ep$ and antipode $S$ 
defined,respectively,by
$$
\begin{array}{l}
\Dl(T(u)_{ab})=\sum _c T(u)_{ac}\otimes T(u)_{cb}\\

\ep (T(u))=1\\

S(T(u))=T(u)^{-1}
\end{array}
\eqno{(3.22a)}
$$
If writting the Hopf structure in terms of operators $\{e_i,f_i,h_i, 
z_i\}_{i=0,1}$,we get the following forms: 
$$
\begin{array}{l}
\Dl(X)=1\otimes X+X \otimes 1\\
\Dl(e_1)=1\otimes e_1 +e_1\otimes 1 -\frac{C_0}{C_1}(h_0 \otimes e_0 + e_0 
\otimes h_0) +\frac{C_0 \ga _0}{2C_1}(z_0 \otimes e_0 - e_0 \otimes z_0) \\
\Dl(f_1)=1\otimes f_1 +f_1\otimes 1 -\frac{C_0}{C_1}(h_0 \otimes f_0 + f_0 
\otimes h_0) +\frac{C_0 \ga _0}{2C_1}(-z_0 \otimes f_0 + f_0 \otimes z_0)\\
\Dl(z_1)=1\otimes z_1 +z_1\otimes 1 +\frac{C_0}{C_1}(e_0 \otimes f_0 - f_0 
\otimes e_0) -\frac{C_0}{C_1}(z_0 \otimes h_0 + h_0 \otimes z_0)\\
\Dl(e
h_1)=1\otimes h_1 +h_1\otimes 1 -\frac{C_0 \ga _0}{2C_1}(f_0 \otimes e_0 + e_0 
\otimes f_0) -\frac{C_0}{C_1}h_0 \otimes h_0 - \frac{C_0 \ga _0^2}{4C_1}z_0 
\otimes z_0 \\
S(X)=-X\\
S(e_1)=-e_1 -\frac{C_0}{C_1}(h_0 e_0 +e_0 h_0)\\
S(f_1)=-f_1 -\frac{C_0}{C_1}(h_0 f_0 +f_0 h_0)\\
S(z_1)=-z_1 -\frac{C_0}{C_1}(f_0 e_0 -e_0 f_0 +2z_0 h_0 )\\
S(h_1)=-h_1 -\frac{C_0 \ga _0}{2C_1}(f_0 e_0 +e_0 f_0 )-\frac{C_0 \ga 
_0^2}{4C_1}z_0 z_0 -\frac{C_0}{C_1}h_0 h_0\\
\ep (1)=1, \ep (X)=\ep (Y)=0
\end{array}
\eqno{(3.22b)}
$$
where $X=e_0,f_0, z_0 ,h_0, Y=e_1, f_1, z_1, h_1 $.

Now we introduce a set of bosonic oscillators $b_i,b^{\dagger}_i$ and a set of 
fermionic oscillators $a_i,a^{\dagger}_i$ satisfying

$$
\left \{
\begin{array}{l}

\{a_i,a^{\dagger}_j \}=[b_i,b^{\dagger}_j ]= \dl _{ij}\\

\{a_i,a_j \} =\{a^{\dagger}_i,a^{\dagger}_j \}=[b_i,b_j 
]=[b^{\dagger}_i,b^{\dagger}_j ]=0\\

[a_i,b_j ]=[a^{\dagger}_i,b^{\dagger}_j ]=[a^{\dagger}_i,b_j 
]=[a_i,b^{\dagger}_j ]=0
\end{array}
\right.
\eqno{(3.23)}
$$
Identifying

$$
\left \{
\begin{array}{ll}
e_0=\sum _{i} b^{\dagger}_i a_i,&f_0 =\sum _{i}a^{\dagger}_i b_i\\

z_0=\sum _{i}(a^{\dagger}_i a_i+b^{\dagger}_i b_i),&h_0=\sum _{i} b^{\dagger}_i 
b_i,\\

e_1=\sum _{i,j} A_{ij}b^{\dagger}_i a_j +\sum _{i,j} B_{ij} b^{\dagger}_i a_i 
(a^{\dagger}_j a_j +b^{\dagger}_j b_j )\\

f_1=\sum _{i,j} A_{ij}a^{\dagger}_i b_j -\sum _{i,j} B_{ij} a^{\dagger}_i b_i 
(a^{\dagger}_j a_j +b^{\dagger}_j b_j )  \\

z_1=\sum _{i,j} A_{ij}(a^{\dagger}_i a_j + b^{\dagger}_i b_j)  \\

h_1=\frac{1}{2} \sum _{i,j} A_{ij}(-a^{\dagger}_i a_j + b^{\dagger}_i b_j)+\sum 
_{i,j} B_{ij} b^{\dagger}_i a_i a^{\dagger}_j b_j  
\end{array}
\right.
\eqno{(3.24)}
$$
where $A_{ij},B_{ij}$ are parameters and $B_{ij}+B_{ji}=0$.We can prove that eqs 
(3.24) reproduce the commutation relations given in eqs (3.18) and (3.19) 
.Substituting  eqs (3.24) into Serre relations (3.21), then there are some 
constrains on $A_{ij},B_{ij}$  and they will be related to parameters $C_0 
,C_1$.

\section {Remarks and Discussions}
\vspace*{-0.35cm}
In this letter ,we only discuss the super-Yangian of the Lie superalgebra 
$gl(1|1)$ and its oscillator realization.The question we should answer is how to 
generalize the discussion to the case of superalgebra $gl(m|n)$ and other 
superalgebras.However,this is connected with physical problems,i.e. wether there 
exist integerable models with $\RMA$ matrix associated with Lie superalgebras. 
As a first step, we wish find a model with the super-Yangian symmetry we have 
discussed.This problem asks for a further study of super-Yangian and its 
representation theory.\\

From section 3, we see that (super-)Yangian is related to the (graded) $RTT$ 
relation.\\Acturally,there are dual relations to the (graded) $RTT$ 
relation,their corresponding algebras is not contained in the (super-)Yangian. 
Yangian double considers all algebraic information contained in $RTT$ relation 
and its dual relations. The Yangian double for simple Lie algebras become an 
interesting research object recently\cite{Kho,Kho/Io}. Naturally,the 
super-Yangian double and the related problems also need to be studied.The work 
in this respect is under investigation.

\end{document}